\documentclass[a4paper,12pt]{article}
\usepackage{fancybox}
\usepackage[dvips]{graphicx}
\usepackage{amsmath}
\usepackage{amsfonts}
\usepackage{mathrsfs}
\usepackage{mathrsfs}
\usepackage{bm}
\usepackage{graphicx}
\usepackage{comment}
\usepackage{multicol}

\usepackage{slashbox}
\usepackage{indent}

\renewcommand{\baselinestretch}{0.8}

\makeatletter 
  \@addtoreset{equation}{section}
 \def\section{\@startsection {section}{1}{\z@}{-3.5ex plus -1ex minus -.2ex}{2.3ex plus .2ex}{\normalsize\bf}}
\makeatother

\begin{document}

\begin{center}
\textbf{\Large
{A rigorous proof of the scallop theorem and a finite mass effect of a microswimmer}}
\end{center}

\begin{center}
Kenta Ishimoto
 and Michio Yamada\\
\textit{
Research Institute for Mathematical Sciences, Kyoto University, Kyoto
 606-8502 Japan}\\

(Dated: July 29, 2011)
\end{center}
~
\renewcommand{\baselinestretch}{1.0}\selectfont
\noindent
We reconsider fluid dynamics for a self-propulsive swimmer in Stokes
 flow. With an exact definition of deformation of a swimmer,
 a proof is given to Purcell's scallop theorem including the body rotation. The
 breakdown of the theorem due to a finite Stokes number is discussed by using a
 perturbation expansion method 
and it is found that the breakdown generally
 occurs at the first order of the Stokes number.
In addition, employing the Purcell's ``scallop'' model, 
we show that the theorem holds up to a higher order if the strokes of the swimmer has some symmetry.\\
~
\renewcommand{\baselinestretch}{1.0}\selectfont
\section{Introduction}
\label{Introduction}

Fluid dynamics of locomotion of microorganisms
 such as bacteria and planktons 
has been studied for more than half a century
(\cite{Lighthill1975},\cite{Taylor1951})
and is still a hot topic in physics, mathematics 
and  biology \cite{Lauga2009}. 
As a milestone to discuss the locomotion of such microswimmers,
there exists a well-known theorem
 called \textit{Purcell's scallop theorem}
 \cite{Purcell1977}, 
which asserts that a microorganism with a {\it reciprocal} stroke in Stokes fluid 
cannot travel at all in one period of its motion.
The proof of the theorem given in Purcell's famous lecture \cite{Purcell1977}
was only schematic. 
Although the theorem were repeatedly discussed 
by many researchers (\cite{Chambrion2010},\cite{Childress1981},\cite{DeSimore2008},\cite{Koiler1996},\cite{Shapere1989})
, the definition of the deformation of the swimmer appears not to have been 
paid much attention.  
Shapere and Wilczek \cite{Shapere1989}
 first established a theoretical formalism for
 a swimmer in viscous fluid in terms of 
gauge structure, and gave a proof for the scallop theorem.
However, their theoretical framework
 based on gauge fields was conceptual and did not
provide operational formulae for the locomotion of the swimmer.
Yariv \cite{Yariv2006} improved
 their framework in order that we can calculate  
  kinematic properties
 like velocity of the swimmer from its surface deformation 
from the fluid dynamical point of view.
However, the scallop theorem was not completely proved in
 his paper, because according to his definition of the deformation 
of the body the rotation 
and the surface deformation of the swimmer are not uniquely identified 
as was pointed out by Yariv \cite{Yariv2006} himself. 
Also Childress and Dudley \cite{Childress2004} states
\begin{quote}
``As far as we know there has been no rigorous proof of this theorem
 based upon the mechanics of a Navier-Stokes fluid and free-swimming body''
\end{quote}
on the present status of the scallop theorem.

In the scallop theorem, on the other hand, the inertia of the fluid and the 
body are totally neglected. 
The scallop theorem and its breakdown have been recently
looked back again \cite{Lauga2011}, and Childress \& Dudley 
\cite{Childress2004} 
discussed the possibility of sudden breakdown of the theorem at a 
nonzero critical Reynolds number. 
The breakdown of the theorem may also arise 
from multiple degrees of freedom of deformation
 and from fluid properties such as rheology \cite{Lauga2009b}
and inertia \cite{Childress2004},\cite{Lauga2007}.
The breakdown due to a finite mass of the swimmer, i.e. a finite
 {\it Stokes number}, was
first discussed by Gonzalez-Rodriguez and Lauga 
\cite{Gonzalez2009}. They provided general differential equations 
that govern locomotion of such a dense swimmer, and suggested that
the scallop theorem does not hold at an arbitrary nonzero Stokes 
number.
However, the definition of the deformation of a swimmer 
is not uniquely defined there, either. 

Here, in this paper, we make clear the definition of the deformation of a
swimmer, and give a rigorous proof for the scallop theorem including rotation
 together with translation. 
We then provide a perturbational argument on the breakdown of the scallop theorem
by a finite inertial effect of a swimmer based on the equations of motion
of both the fluid and the swimmer.
Also with the Purcell's scallop model, we consider a higher order 
breakdown of the scallop theorem under some assumptions on a symmetry of 
the stroke of the swimmer.
For discussion of these problems we develop a theoretical framework 
to describe the motion of microorganisms, 
introducing a virtual swimmer, which undergoes the same surface 
deformation as the real swimmer but without ambient fluid, to 
define rigorously the rotation and the deformation of the real 
swimmer.  We call the coordinates attached to the virtual swimmer 
the \textit{vacuum coordinates}, and the coordinates to the real
swimmer the \textit{body coordinates}. 
The transformation from the former to the latter defines the rotation (i.e. the gauge) of the swimmer.

Some remarks should be made on physical situations 
of a finite Stokes number.
As far as we consider typical microswimmers
 such as bacteria and mammalian sperms, 
 the Reynolds number is small enough for the Stokes equation 
to be available. 
In reality, bodies of the microswimmers are usually a little heavier 
than the surrounding fluid.
Among these microorganisms, relatively larger members
 like \textit{Volvox} and \textit{Paramecium}
may have nonnegligible Stokes numbers, 
$R_S \sim 1$, and smaller Reynolds numbers, 
 $Re\sim 10^{-2}$ and $Re\sim 10^{-1}$ for each sample organism.
A tiny bug in air may be another example for this
situation because the averaged density of such a bug is much 
larger than that of the air.

Here, in this paper, we consider the locomotion of the
microorganisms which have a small but finite Stokes number in the
fluid governed by the steady Stokes equation. 
Of course gravity effects on these swimmers become 
significant at high Stokes numbers, but in this paper
we do not pay much attention to the gravity effects to keep the model 
as simple as possible.
This paper consists of 5 sections.
Section 1 is the introduction, and in 
section 2 we discuss
a theoretical framework for a swimmer 
immersed in Stokes fluid and give an
equation that governs the locomotion of the swimmer. 
In section 3 we restate
 the scallop theorem and
give a complete proof to the theorem.
In section 4, a perturbational discussion of 
the breakdown of the scallop theorem due to a finite Stokes number
 is made for a swimmer without rotation.
In section 5 we consider the breakdown of the scallop theorem
under the assumption of a symmetry of the stroke
using the Purcell's scallop model \cite{Becker2003}.
Summary and conclusion are given in section 6. 

\section{Formulation}
\label{formulation}

\begin{figure*}[!htb]
 \centering
  \includegraphics[width=7cm,clip]{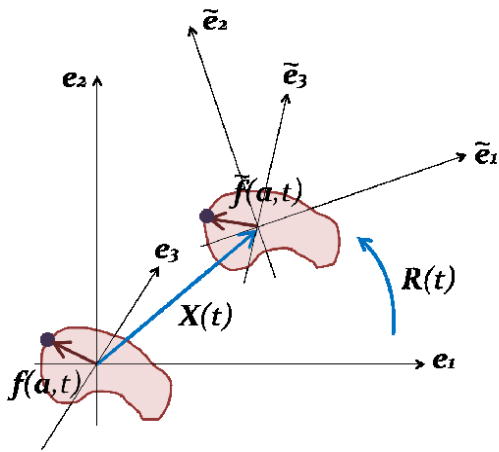}
  \caption{The vacuum coordinates $\{\bm{e}_i\}$ and the body
 coordinates $\{\tilde{\bm{e}}_i\}$.}
   \label{scallop2}
\end{figure*}

In this section we set up formulae for velocity and angular velocity
of a self-propulsive swimmer in fluid.
To discuss the motion of the swimmer, we first define a {\it virtual swimmer}, which
deforms its body in exactly the same way as the real swimmer except that the virtual swimmer has no surrounding fluid, and thus experiences no external forces. 
The virtual swimmer therefore conserves the total momentum and the
total angular momentum both of which we assume to be zero. 
We further assume that at the initial time the virtual swimmer
exactly coincides with the real swimmer, and therefore
both their centers of mass locate at the same position, and their 
orientations are the same. We attach an inertial coordinates, which we call
the \textit{vacuum coordinates}, to this virtual swimmer with its 
origin located at the center of mass of the virtual swimmer. 
The virtual swimmer has exactly the same shape as the real 
swimmer at the same time, and thus the latter is obtained by an affine 
transformation from the former.  We also define the \textit{body coordinates}
attached to the real swimmer as the coordinates obtained by the 
same affine transformation from the vacuum coordinates.
Thus the origin of the body coordinates is located at the center of mass of the real swimmer.
We denote the orthonormal basis of the vacuum coordinates 
by $\bm{e}_i$($i=1,2,3$), 
which is independent of time (Fig. \ref{scallop2}).
The motion of the virtual swimmer is described in Lagrangian coordinates
where a position of a Lagrangian particle of the virtual swimmer,
$\bm{f}(\bm{a},t)=\sum_if_i(\bm{a},t)\bm{e}_i$, is regarded as a function of
the Lagrangian coordinates $\bm{a}=(a_1, a_2, a_3)$ and time $t$
with $\bm{f}(\bm{a},0)=\bm{a}$ .

We now define the surface deformation velocity $\bm{u}'$ of 
the virtual swimmer
as $\bm{u}'= \sum_i \frac{\partial f_i}{\partial t}(\bm{a},t)\bm{e}_i$, 
where $\bm{a}$ is assumed to be on the surface of the swimmer. 
The real swimmer, however, not only deforms
but also translates and rotates under the action of the external 
force from the surrounding fluid.
The translation velocity $\bm{U}$ is defined as $\bm{U}=d\bm{X}/dt$ where
$\bm{X}(t)$ is the center of mass of the real swimmer. 
The position of the Lagrangian particle of the {\it real} swimmer 
$\tilde{\bm{f}}(\bm{a},t)$ with respect to the center of mass of 
the real swimmer is then obtained as 
$\tilde{\bm{f}}(\bm{a},t)=\bm{R}(t)\bm{f}(\bm{a},t)$, 
where $\bm{R}(t)$ is a rotation 
matrix in $SO(3)$ and $\bm{R}(0)=\bm{1}$, 
because the real and virtual swimmers have exactly 
the same shape.  The unit vectors of the body coordinates $\tilde{\bm{e}}_i
 (i=1,2,3)$  is then obtained as $\tilde{\bm{e}}_i(t)=\bm{R}(t)\bm{e}_i$.

The total surface velocity $\bm{u}$ of the real swimmer is
\begin{equation}
\bm{u}= \bm{U}+\cfrac{\partial \tilde{\bm{f}}}{\partial t}={\bm{U}}+
\sum_{i=1}^3\tilde{f}_i\cfrac{d \tilde{\bm{e}}_i}{dt}+\sum_{i=1}^3 
\cfrac{\partial  \tilde{f}_i}{\partial t}\tilde{\bm{e}}_i
=\bm{U}+\bm{\Omega}(t)\times\tilde{\bm{f}}+\tilde{\bm{u}}'
\label{eq6},
\end{equation}
where we have written 
$\tilde{\bm{f}}$ as $\tilde{\bm{f}}=\sum_i \tilde{f}_i \tilde{\bm{e}}_i$, 
and used the fact that $f_i(\bm{a},t)=\tilde{f}_i(\bm{a},t)$ and
$\dot{\tilde{\bm{e}}}_i= \bm{\Omega}\times \tilde{\bm{e}}_i$. The
last term $\tilde{\bm{u}}'$ means the surface deformation velocity of the real
swimmer defined as $\tilde{\bm{u}}'=\sum_i \frac{\partial \tilde{f}_i}{\partial t} \tilde{\bm{e}}_i$ which is equal to $\bm{R}\bm{u}'$. The rotational angular
velocity vector $\bm{\Omega}(t)$ is defined together with two skew symmetric 
matrices $\bm{A}(t), \bm{B}(t)$ as 
\begin{eqnarray}
 \cfrac{d\bm{R}(t)}{dt}&=& \bm{B}(t)\bm{R}(t)=\bm{R}(t)\bm{A}(t)=
\bm{\Omega}\times\bm{R}(t), \label{eq7}\\
\Omega_i(t)&=&-\epsilon_{ijk}B_{jk}(t), \label{eq24}
\end{eqnarray}
where the Einstein convention for repeated indices is employed.  
The angular velocity vector $\bm{\Omega}(t)$ can also be written as 
$\bm{\Omega}(t)=(1/2)\sum_i \tilde{\bm{e}}_i\times \dot{\tilde{\bm{e}}}_i$.

The governing equation of the fluid surrounding the real swimmer
is the incompressible Navier-Stokes equation.
The equation for fluid velocity $\bm{v}=\bm{v}(\bm{x},t)$ is written in a 
non-dimensional form as follows:
\begin{align}
\nabla\cdot\bm{\sigma} &=R_\omega\cfrac{\partial\bm{v}}{\partial
 t}+Re\left(\bm{v}\cdot\nabla\right)\bm{v}
\label{eq1}
\\
\nabla\cdot\bm{v}&=0
\label{eq3}
\end{align}
and the stress tensor $\bm{\sigma}$ is given by
\begin{equation}
\bm{\sigma}=-p\textbf{1}+\left(\nabla\bm{v}+(\nabla\bm{v})^{\text{T}}\right)
\label{eq2}.
\end{equation}
The non-dimensional parameters of the equation, $Re$ and $R_\omega$,
are the Reynolds number $Re=\rho VL/\mu$ and the oscillatory Reynolds
number $R_\omega=\rho L^2\omega/\mu$, the latter of
 which corresponds to the product of the Reynolds number and the
Strouhal number $St=L\omega/V$, where $L$, $V$ and $\omega$ are respectively 
the characteristic scales of length, velocity and frequency ($d/dt$)
of the fluid motion, with the density $\rho$ and the viscosity $\mu$ of the fluid
 being assumed constant in this paper.
The boundary condition for $\bm{v}$ at the body surface of the real swimmer is 
that $\bm{v}$ coincides with the total surface velocity $\bm{u}$ of the real swimmer, while $\bm{v}\rightarrow 0$ at infinity.

From Newton's equation of motion for the real swimmer, we obtain the following
equation in the non-dimensional form,
\begin{equation}
 R_S
\cfrac{d}{dt}
\begin{pmatrix}
\bm{U} \\
\bm{I}\bm{\Omega} 
\end{pmatrix}
=
\begin{pmatrix}
\bm{F} \\
\bm{T} 
\end{pmatrix}
\label{eq4},
\end{equation}
where $\bm{I}$ is the inertial momentum tensor of the real swimmer. 
The non-dimensional number $R_S$, the Stokes number, is
represented as $R_S=\rho_ML^2\omega/\mu$
where $\rho_M$ is the mean density of the swimmer.
The force $\bm{F}$ and the torque $\bm{T}$ acting on the real swimmer 
also non-dimensionalized like 
$\bm{F}^*=\mu LV \bm{F}$ using the Stokes law of resistance, 
where the asterisk denotes the dimensional quantity.
This non-dimensionalization is effective in a situation without
other external forces such as gravity and
electromagnetic forces.

Hereafter, we assume that the non-dimensional parameters $Re$，
$R_\omega$，$R_S$ satisfy the inequality, $Re, R_\omega \ll R_S \ll 1$.
In this section we consider the motion of the swimmer in the fluid with $Re=R_\omega=0$. We will derive equations for the velocity $\bm{U}$ and the angular 
velocity $\bm{\Omega}$. 
When the fluid obeys the steady Stokes equation,   
Lorentz' reciprocal theorem \cite{Lorentz1906} gives
\begin{equation}
 \int_SdS\bm{n}\cdot\bm{\sigma}\hat{\bm{u}}=\int_SdS\bm{n}\cdot\hat{\bm{\sigma}}\bm{u},
\label{eq8}
\end{equation}
where $S$ denotes the surface of the real swimmer, and 
the symbol ``hat'' indicates quantities of another solution of the
Stokes equations with the same boundary shape $S$ and the vector
$\bm{n}$ denotes the unit normal vector to the surface.
For the solution $\hat{\bm{u}}$ we take a solution satisfying the boundary 
condition,
\begin{equation}
\hat{\bm{u}}=\hat{\bm{U}}+\hat{\bm{\Omega}}\times\tilde{\bm{f}} \label{total surface velocity}
\end{equation}
at the surface $S$ of the swimmer \cite{Stone1996}. 
With linearity of the Stokes equation, the stress tensor is given by 
\begin{equation}
 \hat{\bm{\sigma}}=\tilde{\bm{\Sigma}}_T\hat{\bm{U}}+\tilde{\bm{\Sigma}}_R\hat{\bm{\Omega}}
\label{eq9},
\end{equation}
where $\tilde{\bm{\Sigma}}_T$ and $\tilde{\bm{\Sigma}}_R$ are third 
rank tensors \cite{Happel1965} which depend upon the direction 
of the body coordinates $\tilde{\bm{e}}_i$.
Substituting equations (\ref{total surface velocity}) and (\ref{eq9}) into (\ref{eq8}) we have 
\begin{equation}
\begin{pmatrix}
\bm{F}\\
\bm{T}
\end{pmatrix}
\cdot 
 \begin{pmatrix}
\hat{\bm{U}} \\
\hat{\bm{\Omega}}
\end{pmatrix}=
\int_SdS
\begin{pmatrix}
 (\bm{n}\cdot\tilde{\bm{\Sigma}}_T)^{\textrm{T}}\bm{u} \\
 (\bm{n}\cdot\tilde{\bm{\Sigma}}_R)^{\textrm{T}}\bm{u} 
\end{pmatrix}
\cdot
\begin{pmatrix}
\hat{\bm{U}} \\
\hat{\bm{\Omega}}
\end{pmatrix}
\label{eq10},
\end{equation}
where the superscript $\text{T}$ denotes the transpose of a matrix.
Here  $\bm{n}\cdot\tilde{\bm{\Sigma}}_T$ and $\bm{n}\cdot\tilde{\bm{\Sigma}}_R$ denote the second rank
tensors $n_i\tilde{\Sigma}_{Tijk}$ and $n_i\tilde{\Sigma}_{Rijk}$ respectively.
Using the $6\times6$ symmetric resistive matrix
\footnote{
The relation of off-diagonal components of $\mathsf{K}$ is as below using Lorentz' reciprocal theorem.
If we take two solutions with the boundary condition $\bm{u}=\bm{U}$ and $\hat{\bm{u}}=\bm{\Omega}\times\bm{f}$ respectively at the surface $S$, we obtain $\bm{\sigma}=\bm{\Sigma}_T\bm{U}$  and $\hat{\bm{\sigma}}=\bm{\Sigma}_R\bm{\Omega}$ . Substitution of these into equation (\ref{eq8}) gives
$$
\int_SdS\bm{n}\cdot (\bm{\Sigma}_T\bm{U})\bm{\Omega}\times\bm{f}=\int_SdS\bm{n}\cdot(\bm{\Sigma_R\bm{\Omega}})\bm{U}.
$$
After some manipulations, we get
$$
\int_SdS\bm{\Omega}\cdot\bm{f}\times(\bm{n}\cdot\bm{\Sigma}_T)\bm{U}=\int_SdS\bm{\Omega}\cdot(\bm{n}\cdot{\bm{\Sigma}_R})^{\textrm{T}}\bm{U},
$$
which implies that the transpose of an off-diagonal component is another off-diagonal component.
}
\begin{equation}
 \tilde{\mathsf{K}}=
\begin{pmatrix}
\tilde{\bm{K}}_T, & \tilde{\bm{K}}_C^{\text{T}} \\
\tilde{\bm{K}}_C, & \tilde{\bm{K}}_R
\end{pmatrix}
=\int_SdS
\begin{pmatrix}
\bm{n}\cdot\tilde{\bm{\Sigma}}_T, &\bm{n}\cdot\tilde{\bm{\Sigma}}_R\\
\tilde{\bm{f}}\times(\bm{n}\cdot\tilde{\bm{\Sigma}}_T),& \tilde{\bm{f}}\times(\bm{n}\cdot\tilde{\bm{\Sigma}}_R)
\end{pmatrix}
\label{eq11}
\end{equation}
and arbitrariness of $\hat{\bm{U}}$ and $\hat{\bm{\Omega}}$, we obtain
\begin{equation}
 \begin{pmatrix}
\bm{F} \\
\bm{T}
 \end{pmatrix}
=
\tilde{\mathsf{K}}
\begin{pmatrix}
\bm{U}\\
\bm{\Omega}
\end{pmatrix}+
\int_SdS
\begin{pmatrix}
 (\bm{n}\cdot\tilde{\bm{\Sigma}}_T)^{\textrm{T}}\tilde{\bm{u}}' \\
 (\bm{n}\cdot\tilde{\bm{\Sigma}}_R)^{\textrm{T}}\tilde{\bm{u}}'
\end{pmatrix}
\label{eq13}.
\end{equation}
With Newton's equation of motion (\ref{eq4}),
we obtain the desired equation:
\begin{eqnarray}
R_S\cfrac{d}{dt}
 \begin{pmatrix}
\bm{U} \\
\bm{I}\bm{\Omega}
 \end{pmatrix}
&=&
\begin{pmatrix}
\bm{F}_{TR}\\
\bm{T}_{TR}
\end{pmatrix}
+
\begin{pmatrix}
\bm{F}_D\\
\bm{T}_D
\end{pmatrix} \label{EOM}
\\
\begin{pmatrix}
\bm{F}_{TR}\\
\bm{T}_{TR}
\end{pmatrix}
&=&
\tilde{\mathsf{K}}
\begin{pmatrix}
\bm{U}\\
\bm{\Omega}
\end{pmatrix}
\\
\begin{pmatrix}
\bm{F}_D\\
\bm{T}_D
\end{pmatrix}
&=&
\int_SdS
\begin{pmatrix}
 (\bm{n}\cdot\tilde{\bm{\Sigma}}_T)^{\textrm{T}}\tilde{\bm{u}}' \\
 (\bm{n}\cdot\tilde{\bm{\Sigma}}_R)^{\textrm{T}}\tilde{\bm{u}}'
\end{pmatrix}.
\label{eq14}
\end{eqnarray}

The force $\bm{F}_{TR}$ and the torque $\bm{T}_{TR}$ are the external force and the torque from the fluid, arising from the translation and rotation of the real swimmer, while the force $\bm{F}_D$ and the torque $\bm{T}_D$ arise from the surface deformation of the swimmer. 
We remark that the right hand side of equation (\ref{EOM}) represents the total force and torque exerting on the real swimmer, and therefore when we take into account the gravity and the buoyancy effects, we only need to add the gravity and the
buoyancy forces to $\bm{F}$ and their torques to $\bm{T}$, respectively. 
We should note that both the third rank tensors, $\tilde{\bm{\Sigma}}_T$ and $\tilde{\bm{\Sigma}}_R$, and the resistive matrix $\tilde{\mathsf{K}}$ of the real 
swimmer depend only on the surface shape of the real swimmer 
$\tilde{\bm{f}}(t)$.  

As we are considering the scallop theorem which is concerned with the motion of the swimmer due to its surface deformation defined by using the virtual swimmer, it is convenient to describe the problem in terms of quantities of 
the virtual swimmer in the vacuum coordinates.  We then have 
the following equations;
\begin{equation}
R_S\cfrac{d}{dt}
 \begin{pmatrix}
\bm{U} \\
\bm{R}\bm{I}^{V}\bm{R}^{-1}\bm{\Omega}
 \end{pmatrix}
= 
\mathsf{R}\mathsf{K}\mathsf{R}^{-1}
\begin{pmatrix}
\bm{U}\\
\bm{\Omega}
\end{pmatrix}+
\mathsf{R}
\int_SdS
\begin{pmatrix}
 (\bm{n}\cdot\bm{\Sigma}_T)^{\textrm{T}}\bm{u}' \\
 (\bm{n}\cdot\bm{\Sigma}_R)^{\textrm{T}}\bm{u}'
\end{pmatrix},
\label{eq17}
\end{equation}
where the $6\times6$ matrix $\mathsf{R}$ is defined as
\begin{equation}
 \mathsf{R}=
\begin{pmatrix}
\bm{R} & 0 \\
0 & \bm{R} 
\end{pmatrix},
\label{eq18}
\end{equation}
and $\bm{I}^V$ is the inertial momentum tensor of the virtual swimmer. 
When the deformation velocity of the virtual swimmer is given, 
this equation together with (\ref{eq7}) determines the
rotation matrix $\bm{R}(t)$ and the translational velocity $\bm{U}(t)$.

\section{The proof of the scallop theorem}

In this section we give a proof of the scallop theorem using equations (\ref{eq7}) and (\ref{eq17}) 
in the case of the vanishing Stokes number, $R_S=0$. 

We consider the case where the shape of the virtual swimmer deforms in a 
{\it reciprocal} manner, i.e. the shape once deformed retraces back to the 
initial shape.  The mathematical definition of the reciprocal motion is that for the surface deformation 
of the {\it virtual} swimmer, which starts at $t=0$ and ends at $t=T$, 
there exists a continuous function $g(t)$ such that $\bm{f}(t)=\bm{Q}(t)\bm{f}(g(t))$ and $g(0)=g(T)=0$ where $\bm{Q}(t)\in SO(3)$ is a three dimensional rotation matrix, allowing the possibility that the swimmer takes different 
directions at time $t$ and $g(t)$. We assume that $g(t)$ is smooth except 
at a finite number of points, and then intervals of integration over the time $t$ we 
have in the following in this paper should be divided into those 
in which $g(t)$ remains smooth. In this case, we can prove that $\bm{Q}(t)=\bm{1}$ as below.

The total angular momentum of the virtual swimmer always vanishes, and therefore
 \begin{equation}
 \bm{0}=\int \rho_m \bm{f}(\bm{a},t)\times\cfrac{\partial \bm{f}}{\partial t}(\bm{a},t) d\bm{a}=\int \rho_m{\bm{Q}}\bm{f}(\bm{a},g(t))\times\cfrac{\partial{\bm{Q}} \bm{f}(\bm{a},g(t))}{\partial t} d\bm{a}.
 \label{eq19-2}
 \end{equation}
Noting that
\begin{equation}
\cfrac{\partial{\bm{Q}} \bm{f}(\bm{a},g(t))}
{\partial t}=\bm{\Omega}^Q\times \bm{f}(\bm{a},t)+\cfrac{dg}{dt}{\bm{Q}}\cfrac{\partial\bm{f}}{\partial t}(\bm{a},g(t)),
\label{eq19-3}
\end{equation}
where the angular velocity vector $\bm{\Omega}^{Q}$ is defined  as 
\begin{equation}
\frac{d\bm{Q}}{dt}=\bm{\Omega}^Q\times\bm{Q} \label{definition of Q},
\qquad \Omega_i^Q=-\epsilon_{ijk}\left(\frac{d\bm{Q}}{dt}\bm{Q}^{-1}\right)_{jk},
\end{equation}
and $\rho_m=\rho_m(\bm{a})$ is
the density of the swimmer, 
we find the contribution from the second term of equation (\ref{eq19-3}) to (\ref{eq19-2}) vanishes as
\begin{equation}
\int \rho_m{\bm{Q}(t)}\bm{f}(\bm{a},g(t))\times
\cfrac{dg}{dt}{\bm{Q}}\cfrac{\partial\bm{f}}{\partial t}(\bm{a},g(t))\,d\bm{a}
=
\cfrac{dg}{dt}{\bm{Q}(t)}\int \rho_m\bm{f}(\bm{a},g(t))\times
\cfrac{\partial\bm{f}}{\partial t}(\bm{a},g(t))\,d\bm{a}
=0
\end{equation}
due to the vanishing initial angular momentum conserved.
Then the equation (\ref{eq19-3}) gives 
\begin{equation}
\int\rho_m \bm{f}(\bm{a},t)\times \left(\bm{\Omega}^Q(t)\times \bm{f}(\bm{a},t)
\right)\,d\bm{a}=\bm{I}^V(t){\bm{\Omega}^Q}(t)=0 \label{eq19-4}.
\end{equation}
For a general 3-dimensional swimmer, $\bm{I}^V$ is not degenerated and we have $\bm{\Omega}^Q(t)=0$. Then equation (\ref{definition of Q}) with $\bm{Q}(0)=1$ gives $\bm{Q}(t)=\bm{1}$. 

\textit{The scallop theorem} asserts that the position and the direction of the real swimmer at the final time $t=T$ coincide with the initial position and direction if the motion of the swimmer is reciprocal, the surrounding fluid obeys the
steady Stokes equation, and the Stokes number of the swimmer vanishes. 
In this case, the equation (\ref{eq17}) is reduced to
\begin{equation}
 \begin{pmatrix}
\bm{U}\\
\bm{\Omega}
\end{pmatrix}
=-
\mathsf{R}\mathsf{M}
\int_SdS
\begin{pmatrix}
  (\bm{n}\cdot\bm{\Sigma}_T)^{\textrm{T}}\bm{u}' \\
 (\bm{n}\cdot\bm{\Sigma}_R)^{\textrm{T}}\bm{u}'
\end{pmatrix}
\label{eq19},
\end{equation}
where the $6\times 6$ matrix $\mathsf{M}$, the mobility matrix, is the inverse  of the resistive matrix $\mathsf{K}$.

Denoting $t'=g(t)$ for short, the equation (\ref{eq19}) is reduced to 
\begin{align}
 \begin{pmatrix}
\bm{U}(t) \\
\bm{\Omega}(t)
\end{pmatrix}
&=-\mathsf{R}(t)\mathsf{M}(t)
\begin{pmatrix}
\int_{S(t)}dS (\bm{n}\cdot\bm{\Sigma}_T)^{\textrm{T}}(t)\cdot(\partial
 f_i/\partial t)\bm{e}_i \\
\int_{S(t)}dS (\bm{n}\cdot\bm{\Sigma}_R)^{\textrm{T}}(t)\cdot(\partial
 f_i/\partial t)\bm{e}_i  
\end{pmatrix}
\label{eq20}\\
&=-\mathsf{R}(t)\mathsf{M}(t')
\begin{pmatrix}
\int_{S(t')}dS (\bm{n}\cdot\bm{\Sigma}_T)^{\textrm{T}}(t')\cdot(\partial
 f_i/\partial t')\bm{e}_i \\
\int_{S(t')}dS (\bm{n}\cdot\bm{\Sigma}_R)^{\textrm{T}}(t')\cdot(\partial
 f_i/\partial t')\bm{e}_i  
\end{pmatrix}\cfrac{dt'}{dt}
\label{eq21}\\
&=\mathsf{R}(t)\mathsf{R}^{-1}(t')
\begin{pmatrix}
\bm{U}(t') \\
\bm{\Omega}(t')
\end{pmatrix}\cfrac{dt'}{dt}
\label{eq22},
\end{align}
where we have used the fact that $\mathsf{M}(t)=\mathsf{M}(t')$, $S(t)=S(t')$, $\bm{\Sigma}_T(t)=\bm{\Sigma}_T(t')$ and $\bm{\Sigma}_R(t)=\bm{\Sigma}_R(t')$, which holds because $\bm{f}(\bm{a},t)=\bm{f}(\bm{a},t')$. 
Especially from this equation, we obtain 
\begin{equation}
\bm{R}^{-1}(t)\bm{\Omega}(t)=\bm{R}^{-1}(t')\bm{\Omega}(t')\frac{dt'}{dt} .\label{RinverseOmega}
\end{equation}

Using equations (\ref{eq7}) and (\ref{eq24})
we obtain 
\begin{equation}
{A}_{ij}(t)=R_{il}^{-1}B_{lk}R_{kj}=-\epsilon_{lkp}R_{li}R_{kj}\Omega_p ,\label{nanashi2}
\end{equation}
and from the definition of the determinant of the matrix $\bm{R}$ we have
\begin{equation}
 \epsilon_{lkp}R_{li}R_{kj}R_{pq}=|\bm{R}|\epsilon_{ijq}
\label{eq27},
\end{equation}
which leads to 
\begin{equation}
\epsilon_{lkr}R_{li}R_{kj}=\epsilon_{ijq}R_{rq}
=\epsilon_{ijq}(\bm{R}^{-1})_{qr} \label{nanashi}
\end{equation}
by the use of $|\bm{R}|=1$ and the multiplication of $R_{rq}$ to equation (\ref{eq27}).  Substituting equation (\ref{nanashi}) into equation (\ref{nanashi2}), we have 
\begin{equation}
{A}_{ij}(t)=-\epsilon_{ijk}(\bm{R}^{-1}(t) \bm{\Omega}(t))_k ,
\label{eq25}
\end{equation}
which together with (\ref{RinverseOmega}) gives
\begin{equation}
\bm{A}(t)=\bm{A}(t')\cfrac{dt'}{dt}. \label{daiji}
\end{equation}
The rotation matrix $\bm{R}(t)$ satisfies 
\begin{equation}
\frac{d\bm{R}(t)}{dt}=\bm{R}(t)\bm{A}(t). \label{R(t)},
\end{equation}
and $\bm{R}(g(t))$ is also the solution of (\ref{R(t)}), because 
\begin{equation}
\frac{d \bm{R}(g(t))}{dt}= \frac{dg(t)}{dt} \frac{d \bm{R}}{dt}(g(t))
=\frac{dg(t)}{dt} \bm{R}(g(t))\bm{A}(g(t))=\bm{R}(g(t))\bm{A}(t)
\end{equation}
and $\bm{R}(g(0))=\bm{R}(0)=\bm{1}$. Therefore $\bm{R}(g(t))=\bm{R}(t)$, 
and thus $\bm{R}(g(T))=\bm{R}(0)=\bm{1}$\footnote{
This can be proved also by noticing that 
$$
 \bm{R}(T)=\bm{R}(0)\overline{\mathrm{T}}e^{\int_0^{T}\bm{A}(t)dt}=\bm{R}(0)\overline{\mathrm{T}}e^{\int_0^{0}\bm{A}(t')dt'}=\bm{1}.
$$
where $\int_0^0$ symbolize the round integration from $t'=0$ to $t'=0$, and
$\overline{T}$ means anti time-ordering operator.
}. 

Using this relation in equation (\ref{eq22}), we have $\bm{U}(t)=\bm{U}(t')dt'/dt$ which implies
\begin{equation}
 \frac{d \bm{X}(g)}{dt}=\frac{dg(t)}{dt}\frac{d \bm{X}}{dt}(g(t))= 
\frac{dg(t)}{dt}\bm{U}(g(t))=\bm{U}(t).
\end{equation}
Together with $\bm{X}(g(0))=\bm{X}(0)$, this means that $\bm{X}(g(t))=\bm{X}(t)$ and thus $\bm{X}(T)=\bm{X}(g(T))=\bm{X}(0)=\bm{0}$, which completes the proof of the scallop theorem. 

\section{Breakdown of the scallop theorem due to a finite mass of the swimmer}

Let us discuss the breakdown of the scallop theorem by a nonzero Stokes
number. Assuming that the swimmer moves in the fluid with $Re=R_\omega=0$, and
the Stokes number is a small but nonzero constant, we employ $R_S$-expansion as 
$\bm{R},\bm{X},\bm{\Omega}$ and $\bm{U}$:
\begin{align}
 \bm{R}&=\bm{R}_0(1+R_S\bm{R}_1+\cdots) \\
\bm{X}&=\bm{X}^{(0)}+R_S\bm{X}^{(1)}+R_S^2\bm{X}^{(2)}+\cdots \\
\bm{\Omega}&=\bm{\Omega}^{(0)}+R_S\bm{\Omega}^{(1)}+R_S^2\bm{\Omega}^{(2)}+\cdots \\
\bm{U}&=\bm{U}^{(0)}+R_S\bm{U}^{(1)}+R_S^2\bm{U}^{(2)}+\cdots.
\end{align}
At the first order of the Stokes number, substitution of these expansions into 
(\ref{eq7}) and (\ref{eq17}) gives
\begin{equation}
 \bm{\Omega}^{(1)}\times\bm{R}_0=\bm{R}_0\cfrac{d\bm{R}_1}{dt}
\label{eq30_2}.
\end{equation}
and 
\begin{equation}
 \cfrac{d}{dt}
\begin{pmatrix}
\bm{U}^{(0)}\\
\bm{R}_0\bm{I}\bm{R}_0^{-1}\cdot\bm{\Omega}^{(0)}
\end{pmatrix}
=\mathsf{R}_0(\mathsf{R}_1\mathsf{K}-\mathsf{K}\mathsf{R}_1)\mathsf{R}_0^{-1}
\begin{pmatrix}
\bm{U}^{(0)}\\
\bm{\Omega}^{(0)}
\end{pmatrix}
+\mathsf{R}_0\mathsf{K}\mathsf{R}_0^{-1}
\begin{pmatrix}
\bm{U}^{(1)}\\
\bm{\Omega}^{(1)}
\end{pmatrix}
\label{eq30_1} ,
\end{equation}
where the matrices $\mathsf{R}_0$ and $\mathsf{R}_1$ are
$6\times6$ matrices defined as 
\begin{equation}
 \mathsf{R}_0=
\begin{pmatrix}
\bm{R}_0 & 0 \\
0 & \bm{R}_0 
\end{pmatrix},
\qquad
 \mathsf{R}_1=
\begin{pmatrix}
\bm{R}_1 & 0 \\
0 & \bm{R}_1 
\end{pmatrix}.
\end{equation}

Here we focus our attention on a swimmer without rotational motion ($\bm{\Omega}(t)=\bm{0}$). 
Then we can assume that the motion of the swimmer is in the direction of $\bm{e}_1$,  reducing the equation (\ref{eq17}) into
\begin{equation}
 \varepsilon\cfrac{dU(t)}{dt}=-K(t)U+F(t)
\label{eq32},
\end{equation}
where we denote $\varepsilon=R_S$, $U(t)=\bm{U}(t)\cdot \bm{e}_1$, 
$K(t)=-(\bm{K}_T(t)\bm{e}_1)\cdot\bm{e_1}$,
and $F(t)=\bm{F}_D(t)\cdot\bm{e}_1$, and $K(t)$ is positive definite.
Then the equation (\ref{eq32}) is exactly solved as
\begin{equation}
 U(t)=\frac{1}{\varepsilon}e^{-\frac{1}{\varepsilon}\int_0^{t}K(t')dt'}\int_0^tdt'F(t')e^{\frac{1}{\varepsilon}\int_0^{t'}K(t'')dt''}+U(0)e^{-\frac{1}{\varepsilon}\int_0^{t}K(t')dt'}
\label{eq33}.
\end{equation}

Integrating by parts repeatedly\footnote{Denoting $E(t)=
e^{\frac{1}{\varepsilon}\int_0^{t}K(t')dt'}$, we have
$E(t)=\epsilon K^{-1}(t)\dot{E}(t)$ and the integration becomes
$$\int_0^t F(t')E(t')\,dt' 
=\varepsilon \left[E(t')F(t')K^{-1}(t')\right]_0^t- \varepsilon \int_0^t \frac{d}{dt'}\left( F(t')K^{-1}(t')\right) E(t')\,dt'$$ and similarly. },
we represent the equation (\ref{eq33})
 as the form of the asymptotic power series
of the Stokes number $\varepsilon$
 as $U(t)\sim \sum_{n=0}^{\infty}\varepsilon^n U^{(n)}(t)$, 
where we use an assumption that $K(t)$ and $F(t)$ are smooth functions,
and eliminate exponentially small terms from the asymptotic expansion. 
Then the first few terms of the expansion 
are 

\begin{align}
 U^{(0)}(t)&=F(t)K^{-1}(t)
\label{eq37}\\
 U^{(1)}(t)&=-\cfrac{d}{dt}\left(F(t)K^{-1}(t)\right)K^{-1}(t)
\label{eq38}\\
 U^{(2)}(t)&=\cfrac{d}{dt}\left(\cfrac{d}{dt}\left(F(t)K^{-1}(t)\right)K^{-1}(t)\right)K^{-1}(t)\label{eq39}.
\end{align}

We now consider the breakdown of the scallop theorem.
Assume a swimmer in a reciprocal motion. 
Since the propulsive force $F$ satisfies $F(t)=F(t')\frac{dt'}{dt}$ 
and $K(t)=K(t')$, where $t'=g(t)$, the zeroth order of 
equation (\ref{eq32}) gives
\begin{equation}
X^{(0)}(T)=\int_0^T F(t)K^{-1}(t)\,dt =\int_0^0 F(t')K^{-1}(t')\,dt'=0 ,
\end{equation}
which corresponds to the scallop theorem\footnote
{Here again we have used the symbolic notation $\int_0^0$.}．
According to equation (\ref{eq38}),
the first-order displacement in one period of the reciprocal motion
 is represented as
\begin{equation}
 X^{(1)}=\int_0^TdtU(t)=-\int_0^{T}dt\cfrac{dU^{(0)}(t)}{dt}K^{-1}(t). \label{eqX1}
\end{equation}
In the case of the reciprocal motion
as $g(t)=t$ on $[0,T_r]$ and $g(t)=(T-t)T_r/(T-T_r)$
on $[T_r,T]$\footnote{The $dg(t)/dt$ is discontinuous at $t=T/2$, and we divide the integral interval into $[0,T_r]$ and $[T_r,T]$}, equation (\ref{eqX1}) leads to 
\begin{align}
 X^{(1)}(T)=&\frac{T}{T-T_r}\int_0^{T_r}dt U^{(0)}(t)\cfrac{dK^{-1}(t)}{dt} \nonumber \\
 &+\frac{T}{T-T_r}\left(U^{(0)}(+0)K^{-1}(+0)-U^{(0)}(T_r-0)K^{-1}(T_r-0)\right).
\end{align}
This does not necessarily vanish as we show in the next section
That $X^{(1)}$ does not generally vanish, means that the
scallop theorem breaks down at the first order of the Stokes number.
This results are consistent with the continuous breakdown of the scallop
theorem suggested by Gonzalez-Rodriguez and Lauga \cite{Gonzalez2009}.
If, however, the stroke of the swimmer has some symmetry, the 
first-order displacement in one period of the stroke can also vanish, and
the theorem holds true up to a higher order of the Stokes number, as also
discussed in the next section.

\section{Purcell's scallop model}
\label{model}

\begin{figure}[!htb]
 \centering
  \includegraphics[width=6cm,clip]{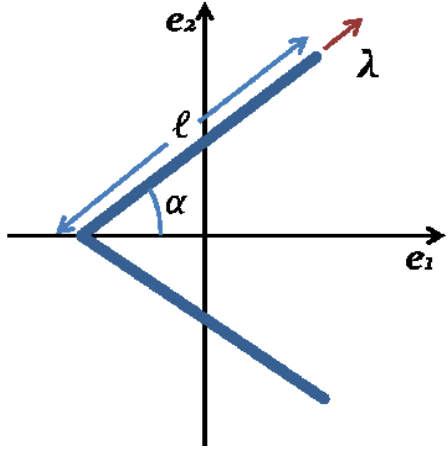}
  \caption{The conceptual figure of the virtual swimmer of the Purcell's ``scallop''．
The coordinates $\bm{e}_1,\bm{e}_2$ are the vacuum coordinates with the origin at the center of mass of the \textit{virtual} ``scallop''. 
}
   \label{scallop1}
\end{figure}

In this section we discuss the motion of the Purcell's ``scallop'' model
\cite{Becker2003} as an example of a swimmer with a finite Stokes
number (Fig. \ref{scallop1}).
This swimmer with one hinge to move
 was first introduced by Purcell to
explain the scallop theorem  \cite{Purcell1977}.
The ``scallop'' has two slender rods with the same length $l$ and with the same
circular cross section of radius $r$. From the \textit{symmetry} of the shape, 
it does not rotate and so become an example of the swimmer without rotation．
We calculate resistive and propulsive forces 
using Cox's slender-body theory \cite{Cox1970},
which describes approximately well when
the aspect ratio of the body $a=r/l$
 is much smaller than 1.
Let us take a coordinate $s$（$-l<s<l$）along the body
from the bottom edge to the top, with the point of $s=0$ at the 
hinge. Let $\bm{\lambda}(s)$ denote the normal vector to the cross section．
Cox's theory gives the force exerted by fluid on an infinitesimal part 
of the body between $s$ and $s+ds$ as
\begin{equation}
 d\bm{F}(s)= \left(\int_{L(s)
  }\bm{n}\cdot\bm{\sigma}dl \right)\,ds =-\zeta(2-\bm{\lambda}\bm{\lambda})\cdot\bm{u}\,ds
\label{eq43},
\end{equation}
where $\zeta=2\pi\mu(\log{(2/a)})^{-1}$.  Here the line integral is taken along the 
circumference $L(s)$ of the cross section at the position $s$, 
and $\bm{u}(s)$ indicates the total surface velocity of the part of the body, 
which is equivalent to the fluid velocity at the surface. 

\begin{figure}[!htb]
 \centering
  \includegraphics[width=12cm,clip]{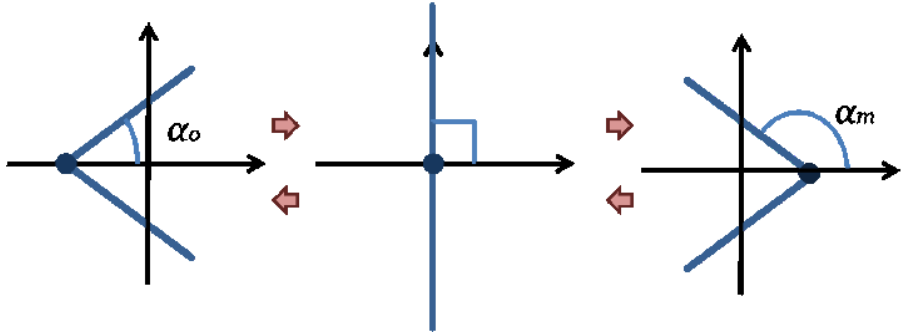}
  \caption{Temporal variation of the shape of the swimmer.}
   \label{scallop4}
\end{figure}

The scallop is configured in two dimensional $\bm{e}_1\bm{e}_2$-plane as shown in Fig. \ref{scallop1}. The motion is then in one-dimensional direction. 
The motion is driven by temporal variation of the angle $\alpha$; $\alpha(t)$ changes reciprocally as $\alpha_0\rightarrow \alpha_m\rightarrow \alpha_0$ (Fig. \ref{scallop4}).
The slender-body theory gives $\int_{L(s)}\,dl(\bm{n}\cdot\bm{\Sigma}_T)^{\textrm{T}}$ and the resistive coefficient $K$ as
\begin{eqnarray}
\int_{L(s)}\,dl(\bm{n}\cdot\bm{\Sigma}_T)^{\textrm{T}}(s)
&=& -\zeta\begin{pmatrix}
2-\cos^2\alpha & -\textrm{sgn}(s)\cos\alpha\sin\alpha \\
-\textrm{sgn}(s)\cos\alpha\sin\alpha & 2-\sin^2\alpha
\end{pmatrix}
\label{eq45}, \\
 K&=&2\zeta l(2-\cos^2\alpha)
\label{eq44},
\end{eqnarray}
in the vacuum coordinates.
When we write the position and the velocity of the hinge as
$X_H(t)=-0.5l\cos\alpha$ and
$U_H(t)= +0.5l\sin\alpha (d\alpha/dt)$, 
it is found that the surface deformation velocity of the virtual swimmer is
$\bm{u}'(s)=(U_H,0)+s(d\bm{\lambda}/dt)$,
where $\alpha=\alpha(t)$ depends
on time $t$ and hereafter we assume some smoothness of $\alpha(t)$ so that $\dot{\alpha}(0)=\ddot{\alpha}(0)=0$. 
The propulsive force $F$ is then derived from (\ref{eq45}) as
\begin{equation}
 F=2\zeta l^2 \sin\alpha\cos^2\alpha\cfrac{d\alpha}{dt}
\label{eq46}.
\end{equation}

Substitution of (\ref{eq44}) and (\ref{eq46}) into equation (\ref{eq32})
gives
\begin{equation}
 \varepsilon'\cfrac{dU}{dt}=-(2-\cos^2\alpha)U+\cos^2\alpha \sin\alpha\cfrac{d\alpha}{dt}l
\label{eq47},
\end{equation}
where $\varepsilon'$ is defined as $\varepsilon'=a^2\log(2/a)\varepsilon/2\omega$,
and $\omega$ is the frequency of the stroke of the swimmer.
Henceforth we nondimensionalize quantities, using $l$ and $1/\omega$ for
the unit of length and time. 
Let us assume $\varepsilon'\ll 1$ and consider the expansion of
 the equation (\ref{eq47}) in terms of $\varepsilon'$．
At the zeroth order of $\varepsilon'$，we get the velocity $U^{(0)}$ as
\begin{equation}
 U^{(0)}(t)=\cfrac{\cos^2\alpha\sin\alpha}{2-\cos^2\alpha}\left(\cfrac{d\alpha}{dt}\right)
\label{eq48}
\end{equation}
and thus the displacement of the center of mass in one period becomes
\begin{equation}
 X^{(0)}(t)=\int_0^t
  U^{(0)}(t')\,dt'=\int_{\alpha_0}^{\alpha}\cfrac{\cos^2\alpha'\sin\alpha'}{2-\cos^2\alpha'}d\alpha'
\label{eq49}.
\end{equation}
Since $X^{(0)}$ is a function of $\alpha$,
 the realization of the scallop theorem is confirmed.
At the first order of $\varepsilon'$, equation (\ref{eq38}) gives the first order
displacement as
\begin{equation}
 X^{(1)}(T)=\int_0^{T}U^{(1)}(t)dt=-\int_0^{T}\cfrac{2\sin^2\alpha\cos^3\alpha}{(2-\cos^2\alpha)^2}\left(\cfrac{d\alpha}{dt}\right)^2dt
\label{eq50}.
\end{equation}
Here we have used integration by parts and $\dot{\alpha}(0)=0$.

We now assume a reciprocal motion in which the swimmer opens and then closes
its ``shell'' during one period.
If $0<\alpha_0<\alpha<\alpha_m<\pi/2$, then
 the integrand of equation (\ref{eq50}) is always positive, and 
the swimmer moves to
$-\bm{e}_1$ direction by the effect of the mass inertia of the swimmer. 

If $0<\alpha_0<\alpha<\alpha_m<\pi$ and $\alpha_m=\pi-\alpha_0$,
the period of motion of the swimmer can then be divided
into 4 intervals,$I_{\alpha_0\rightarrow \pi/2}, I_{\pi/2\rightarrow \alpha_m},I_{\alpha_m\rightarrow \pi/2},I_{\pi/2\rightarrow \alpha_0}$, in each of which the motion is time-reversal or 
mirror-symmetric to that in another interval. 
If the speed of the deformation in $I_{\pi/2 \rightarrow \alpha_m}$ 
is $c_2$ times faster than that in $I_{\alpha_0\rightarrow \pi/2}$,
the first-order displacement in the first half, $ X^{(1)}_{\alpha_0\rightarrow\alpha_m}$, becomes
\begin{equation}
 X^{(1)}_{\alpha_0\rightarrow\alpha_m}=(1-c_2)
  X^{(1)}_{\alpha_0\rightarrow\pi/2}
\label{eq52},
\end{equation}
where $X^{(1)}_{\alpha_0\rightarrow\pi/2}$ denotes the first-order displacement in $I_{\alpha_0 \rightarrow \pi/2}$.
Similarly, if the deformation speeds of in $I_{\alpha_m\rightarrow \pi/2}$ and
$I_{\pi/2\rightarrow\alpha_0}$ are by $c_3$ and $c_4$ times faster than that in $I_{\alpha_0 \rightarrow\pi/2}$ respectively, the net displacement in one period
of the reciprocal motion is
\begin{equation}
 X^{(1)}(T)=(1-c_2-c_3+c_4)X^{(1)}_{\alpha_0\rightarrow\pi/2}
\label{eq53}.
\end{equation}
If $1-c_2-c_3+c_4=0$, we obtain
 $X^{(1)}(T)=0$ and 
the scallop theorem still holds up to the first
 order of Stokes number.

In the second order of $\varepsilon'$, we have
\begin{equation}
 X^{(2)}(T)=-\int_0^T \cfrac{1}{2-\cos^2\alpha}\left(\cfrac{dU^{(1)}}{dt}\right)dt 
=\int_0^T\cfrac{4\sin^3\alpha\cos^4\alpha}{(2-\cos^2\alpha)^5}\left(\cfrac{d\alpha}{dt}\right)^3  dt 
\label{eq54},
\end{equation}
using integration by parts and the assumption that the angle
$\alpha=\alpha(t)$ is a smooth enough function of time
 to satisfy $\ddot{\alpha}(0)=0$.
With the same assumptions on the speed of the motion as in equation (\ref{eq53}), the second-order
displacement is obtained as,
\begin{equation}
 X^{(2)}(T)=(1+c_2^2-c_3^2-c_4^2)X^{(2)}_{\alpha_0\rightarrow\pi/2}
\label{eq55}. 
\end{equation}
Thus the symmetric deformation of $(c_2, c_3, c_4)=(1, 1, 1)$
gives the zero net displacement up to the second order: $X^{(1)}(T)=X^{(2)}(T)=0$. 
On the other hand, the different deformation speeds as
$(c_2, c_3, c_4)=(1,2,2)$ produces the net displacement $X^{(2)}(T)\neq 0$ in
one period of the motion while the first order displacement then vanishes: $X^{(1)}(T)=0$.
This result implies that if the ``shell'' opens slowly and closes quickly, 
the swimmer with the symmetric stroke shown in
Fig. \ref{scallop4} gives the net displacement in $-\bm{e}_1$ direction at
the second order of Stokes number.

Before ending this section, let us consider the case of 
$\varepsilon\gg 1$.
Using equation (\ref{eq33}),
 we expand the velocity of the swimmer
 in Taylor series of $1/\varepsilon$:
\begin{equation}
U(t)=(1/\varepsilon)\int_0^t dt'F(t')+(1/\varepsilon^{2})\int_0^tdt'F(t')\int_t^{t'}K(t'')dt''+\mathcal{O}(1/\varepsilon^{3})
\label{eq56}.
\end{equation}
Applying this equation to equation (\ref{eq47}), we obtain  at the first order of $1/\varepsilon'$
\begin{equation}
U^{(1)}(t)=
\int_{\alpha_0}^{\alpha}\cos^2\alpha\sin\alpha\,d\alpha 
\label{eq56-2},
\end{equation}
which is independent of the speed of deformation.
In the case of $0<\alpha_0<\alpha<\alpha_m<\pi/2$, $X^{(1)}$ is positive quantity.
Thus the ``scallop'' goes to $+\bm{e}_1$ direction when its ``shell'' opens slowly and
 closes quickly as real scallops do.

\section{Summary and Conclusion}

We have established a framework to discuss a motion of a swimmer 
immersed in Stokes fluid, and given a rigorous proof for the scallop theorem.
We have also discussed the breakdown of the theorem due to a nonzero mass of the swimmer, 
and shown that the degree of the breakdown depends on the symmetry of the
 stroke using the Purcell's scallop model.

First of all, in order to define the deformation velocity we introduced the \textit{virtual} swimmer which has the same shape as the \textit{real} swimmer but has no ambient fluid.
We then attached the  \textit{the vacuum coordinates} to the virtual swimmer and \textit{the body coordinates} to the real swimmer.
The position and orientation of the real swimmer is obtained by the affine transformation from the virtual swimmer.
We derived the formulae which provide the velocity and the angular velocity of the swimmer 
when the surface deformation of the virtual swimmer is given 
in the case of vanishing Reynolds number $Re$ and oscillatory Reynolds number $R_\omega$.
Using these formulae, we have proved \textit{the scallop theorem} at a vanishing mass  of the swimmer, or equivalently a zero Stokes number.

Then we studied the breakdown of the scallop theorem, taking a finite Stokes number into consideration especially in the case of the swimmer without rotating motion. 
We showed the net displacement is generally at the first order of the Stokes number
using an asymptotic expansion.
We took Purcell's scallop model with a finite mass as an example and demonstrated the breakdown of the scallop theorem, but also showed that the theorem holds up to a higher order of the Stokes number for a swimmer deforming with a particular symmetry. 

Our argument is based on the assumption 
of $Re, R_\omega \ll 1$ and $Re, R_\omega \ll R_S$.
In reality, however, the time derivative term of the fluid equation may become important for such a dense swimmer,
where $Re \ll R_\omega , R_S$, and $Re \ll 1$ are satisfied.
Gavze \cite{Gavze1990} gave formulae for the fluid dynamical 
force and torque on a rigid (non-deformable) body with an 
arbitrary shape under such conditions.
However, to our knowledge, an explicit form of fluid dynamical force on
self-propulsive deformable swimmer has not yet been found in non-stationary Stokes flow. 
It is still an open question whether the scallop theorem
holds in the case of a finite oscillatory Reynolds number.

The gravity effect on a swimmer may also be important. 
In the case of \textit{Volvox}, the mass density of the body 
is a little heavier than the surrounding water.
According to Drescher et al.\cite{Drescher2009}, 
the density difference between the {\it Volvox} and water is approximately 
$\Delta\rho\sim 2\times 10^{-3}$g/cm$^3$.
The characteristic velocity due to the gravity effect is estimated to be  $U_g\sim 2\times 10^2$cm/sec by balancing the gravity effect with the resistive force on the body,
\begin{equation*}
 \cfrac{\frac{4}{3}\pi A^3 \Delta\rho g}{6\pi \mu A U_g}\sim 1
\label{eq63}.
\end{equation*}
This velocity is comparable to the propulsive velocity
 of the swimmer. 
Burton et al. \cite{Burton2010} has recently
 studied a neutrally buoyant Purcell's scallop model 
with separated positions of the centers of mass 
and of buoyancy.
This kind of separation often occurs in microorganisms
due to heterogeneity of mass distribution, and it may be
of interest to apply our formulation to these organisms
by including gravity and buoyancy. 

Before ending, we should remark the possibility
 of another choice of the gauge fixing to
define the deformation velocity of a swimmer.
In this paper we have introduced a virtual swimmer and its associated
vacuum coordinates to define the deformation velocity.  This choice 
appears natural but is not unique, and 
there may be another coordinate system,
which is useful  in considering other subjects beyond the scallop theorem.

\section*{Acknowledgment}
We would like to thank Prof.\ Takehiro for providing us with
carefully considered feedback and valuable comments. Special thanks are also
to Mr.\ Inubushi, Mr.\ Kimura, Ms.\ Obuse and Mr.\ Sasaki who gave us 
invaluable comments and warm encouragements.

\renewcommand{\baselinestretch}{0}\selectfont

~


\end{document}